\documentclass[prb,twocolumn,amsmath,amssymb,floatfix]{revtex4}
\usepackage{graphicx}
\usepackage{dcolumn} 
\usepackage{bm}
\usepackage{amsmath}
\usepackage{latexsym}
\usepackage{amsfonts}
\usepackage{dcolumn}
\usepackage{color}
\usepackage{amssymb}
\usepackage[rightcaption]{sidecap}
\usepackage{multirow}
\newcommand{\be}{\begin{equation}}
\newcommand{\ee}{\end{equation}}
\newcommand{\bea}{\begin{eqnarray}}
\newcommand{\eea}{\end{eqnarray}}
\newcommand{\ba}{\begin{align}}
\newcommand{\ea}{\end{align}}
\newcommand{\nn} {\nonumber}
\renewcommand{\vr} {{\bf r}}

\newcommand{\Tr}{ {\rm Tr} \, }

\def\a{\alpha}
\def\b{\beta}
\def\g{\gamma}

\def\d{\delta}

\def\ve{\varepsilon}

\def\l{\lambda}
\def\L{\Lambda}

\def\S{\Sigma}

\def\vf{\varphi}

\def\w{\omega}

\def\bra{\langle}
\def\ket{\rangle}

\def\xc{{\rm xc}}

\def\Tr{{\rm Tr}\,}

\def\br{\mbox{\boldmath $r$}}

\begin{document}
%------------------------------------------------------------------------------------------------------------
%--------------------------------------------------Title----------------------------------------------------
%------------------------------------------------------------------------------------------------------------

\title{Correlation potentials for molecular bond dissociation within the self-consistent random phase approximation}
\author{Maria Hellgren$^{1}$}
\author{Daniel R. Rohr$^{1,2}$}
\author{E. K. U. Gross$^{1}$}
\affiliation{$^{1}$Max-Planck-Institut f\"ur 
Mikrostrukturphysik, Weinberg 2, 06120 Halle (Saale), Germany\\
$^{2}$Fritz-Haber-Institut der Max-Planck-Gesellschaft,
Faradayweg 2-4,
14195 Berlin, Germany}
\date{\today}

%------------------------------------------------------------------------------------------------------------
%-----------------------------------------------Abstract-----------------------------------------------------
%------------------------------------------------------------------------------------------------------------
\begin{abstract}
Self-consistent correlation potentials for H$_2$ and LiH for various inter-atomic separations are obtained within the random phase approximation (RPA) of density functional theory. The RPA correlation potential shows a peak at the bond midpoint, which is an exact feature of the true correlation potential, but lacks another exact feature: the step important to preserve integer charge on the atomic fragments in the dissociation limit. An analysis of the RPA energy functional in terms of fractional charge is given which confirms these observations. We find that the RPA misses the derivative discontinuity at odd integer particle numbers but explicitly eliminates the fractional spin error in the exact-exchange functional. The latter finding explains the accurate total energy in the dissociation limit.
\end{abstract}
%\pacs{}
\maketitle
%------------------------------------------------------------------------------------------------------------ 
%---------------------------------------------Introduction---------------------------------------------------
%------------------------------------------------------------------------------------------------------------
\section{Introduction}
The random phase approximation (RPA) in Kohn-Sham (KS) density functional theory (DFT) has in recent years received considerable attention in quantum chemistry\cite{ff01,ff08,eyf10} and material science.\cite{shsgmmk,lhgakd,hk09,rtrs,rrs09,godbyGW1,godbyGW2,gruning} The RPA incorporates exchange effects exactly and the correlation energy is treated non-perturbatively by summing a subset of polarization diagrams to infinite order.\cite{fw,amt02} Furthermore, the RPA can be systematically improved being the first approximation within the so-called adiabatic connection fluctuation dissipation (ACFD) framework.\cite{fg02,kmm98,tgijsaj09}

The RPA is an implicit functional of the density and can therefore include non-local correlation effects like e.g. the van der Waals interactions. That these are, indeed, accurately captured by the RPA has been demonstrated in many recent works.\cite{dobsonrpa,ng10,lhgakd10} For systems described by strong Hubbard-like correlations, the RPA is, however, still not fully investigated. A popular test case in this regard is the dissociation of diatomic molecules with covalent bonds.\cite{htr09} All density functional approximations constructed so far fail in this context if proper spin-symmetry is enforced. The total energy in the dissociation limit is too high and spurious fractional charges are found at the fragments.
 
The large error in the total energy has been characterized as static correlation error or so-called fractional spin error, studied in detail in the pioneering works of Cohen, Mori-S\`anchez and Yang.\cite{fscmy08,mcy09,cmsy08} It has been demonstrated that the RPA strongly improves the dissociation limit for homoatomic systems such as the H$_2$ molecule.\cite{nedvlvb2,fngb05,nedvlvb,hessg11} 

Spurious fractional charges, on the other hand, appear only in the dissociation limit of heteroatomic molecules and are related to an incorrect behavior of the total energy as a function of particle number. The exact energy functional exhibits a kink or derivative discontinuity at integer particle numbers along with a straight line behavior between the integers.\cite{pplb82} The smooth and non-linear behavior of approximate functionals leads to charged fragments in the dissociation limit, meaning that one electron is too delocalized, i.e., spread over both fragments. This error is known in the literature as delocalization error or fractional charge error.\cite{fccmy08,mcy09,cmsy08} The delocalization error of the RPA has been studied only for the dissociation of open-shell H$_2^+$ and He$_2^+$.\cite{yangrpa} It was found to be rather severe leading to too low total energies. Whether the RPA suffers from fractional charge error in the cases of heteronuclear molecules is presently unknown. 

The exact functional ensures neutral dissociation fragments by virtue of the KS potential.\cite{gb96,sgb98,tmm09} In the dissociation limit the highest occupied orbital of each of the fragments must be aligned (or degenerate). Consequently, the highest occupied molecular orbital (HOMO) of the whole system (including both fragments) is a linear combination of the orbitals of each fragment. To obtain equal weights, i.e. integer charge, at the fragments the KS potential exhibits a sharp step at the bond midpoint, shifting the energy levels of only one of the fragments. This feature is a direct consequence of the derivative discontinuity in the correlation part of the energy.\cite{perde,sagper} 
Another feature of the exact KS correlation potential of dissociated molecules is a peak at the bond midpoint.\cite{buijse} The peak emerges with increasing inter atomic distance, and acts to further localize the electrons.

The RPA correlation potential for atoms was recently obtained\cite{hvb1,hvb4} and showed a close resemblance to the exact correlation potential.  However, all RPA calculations on molecules have so far been carried out using potentials originating from other functionals and hence precluding a full assessment of the RPA. The aim of this work is to provide a more complete analysis of the RPA. To this purpose, we have calculated self-consistent RPA potentials for molecules and investigated the RPA correlation potentials for H$_{2}$ and LiH at different inter-atomic distances. These systems allow us to study both the static correlation error and the delocalization error. Moreover, the LiH is an example where a self-consistent calculation is essential. We have also analyzed the RPA energy functional in terms of fractional charge by studying the RPA functional on an extended domain of spin-compensated densities allowing for non-integer number of particles. 

We conclude that the RPA potentials exhibit the peak at the bond midpoint but lack the step feature, where the latter is related to a missing intra-shell derivative discontinuity. The total energy of LiH is, however, still largely improved in the dissociation limit as compared to, e.g., the exact-exchange (EXX) functional, suggesting only a smaller delocalization error.
\section{RPA correlation energy and potential}
\label{RPAenergy}
Within the ACFD framework the exact correlation energy is written as\cite{lp75,gl76,tddftbook}
\be
E_{\rm c}=\frac{i}{2}\int_0^1 d\l\int_{-\infty}^{\infty}\frac{d \w}{2\pi}\,\Tr \{v[\chi_{\l}(\w)-\chi_s(\w)]\}
\label{corr}
\ee
where $\chi_s$ is the non-interacting KS density response function and $\chi_{\l}$ is the scaled interacting density response function. The scaling refers to a system with a linearly scaled Coulomb interaction $\l v(\vr,\vr')$ plus a fictitious potential which keeps the density fixed as $\l$ is changed. The parameter $\l$ runs between 0 (non-interacting KS case) and 1 (fully interacting case). We have used the short hand notation $\Tr fg=\int d{\vr}d{\vr'} f({\vr},{\vr'})g({\vr'},{\vr})$ for any two-point functions $f$ and $g$. 
Within TDDFT the function $\chi^{\l}$ reads\cite{grosskohn} 
\be
\chi_{\l}=\chi_s+\chi_s\left[\l v+f^{\l}_\xc\right]\chi_{\l}.
\label{rpafx}
\ee
The scaled XC kernel $f^\l_{\xc}$ is defined as the functional derivative of the scaled XC potential $v^\l_{\xc}$ with respect to the density $n$, evaluated at the ground state density. 

In the RPA $f^\l_{\xc}=0$ and thus corresponds to the simplest approximation within the ACFD formalism. Within the RPA the $\l$-integral in Eq. (\ref{corr}) can be evaluated analytically with the result
\be
E_{\rm c}=-\frac{i}{2} \int \frac{d\omega}{2\pi}\,\,\Tr\{\ln[1-v\chi_s]+v\chi_s\}.
\label{ec}
\ee
Diagrammatically Eq. (\ref{ec}) is equal to an infinite summation of ring-diagrams.

The RPA correlation potential $v_{\rm c}$ can be obtained as the functional derivative of Eq. (\ref{ec}) 
with respect to the density. If we let $V_s$ signify the total KS potential,  $G_s$ the non-interacting KS Green function and $\chi_s=-iG_sG_s$, the functional derivative can be obtained via the chain rule 
\be
n_c\equiv\frac{\d E_{\rm c}}{\d n}\frac{\d n}{\d V}=\frac{\d E_{\rm c}}{\d G_s}{\frac{\d G_s}{\d V}}.
\label{derivative}
\ee
The result is the well-known linearized Sham-Schl\"uter (LSS) equation\cite{ss83,vbdvls05}
\be
\int\chi_s(1,2)v_{\rm c}(2)d2=\int \L(3,2;1)\S_{\rm c}(2,3)d2d3.
\label{lss}
\ee
Here, we have used the notation $(\vr_1,t_1)= 1$ etc. and introduced
$\L(3,2;1)=-iG_s(3,1)G_s(1,2)$. The correlation part of the self-energy $\S_{\rm c}$ in the RPA 
is given by
\be
\S_{\rm c}=i\frac{\delta E_{\rm c}}{\delta G_s}=iv\chi^{\rm RPA} vG_s
\label{sigma}
\ee
where 
\be
\chi^{\rm RPA}=\chi_s+\chi_sv\chi^{\rm RPA}.
\label{rpa}
\ee
In the appendix we show the expression for $n_c$, i.e., the right hand side of Eq. (\ref{lss}), in terms of KS orbitals and KS eigenvalues.
\section{Fractional charge and spin}
\label{RPAdissociation}
The RPA functional produces accurate dissociation energies for H$_2$, in contrast to all common density functionals which yield a far too high energy due to a spurious self-interaction in the H fragments. It is well-known that EXX is self-interaction free in the case of a spin-polarized H atom. In the dissociation limit of H$_2$, the H atoms are, however, not spin-polarized,
but rather described by a mixture of a spin-up and a spin-down H atom, in which case EXX does suffer from self-interaction. To obtain the correct dissociation limit a significant correlation contribution is thus needed. In the following we will show that the RPA correlation functional exactly cancels the spurious self-interaction in the EXX functional in the dissociation limit. The total energy will still not be exact as the correlation energy contains a self-correlation term, which does not vanish in the dissociation limit.

\subsection{RPA in the dissociation limit}
The RPA energy is an explicit functional of the KS Green function $G_s$ and for spin-compensated systems $G_s$ is proportional to the identity matrix in spin-space. In the frequency domain a spin-component of $G_s$ reads 
\be
G_s(\vr,\vr',\w)=\sum_k^{\rm occ}G^-_k(\vr,\vr',\w)+\sum_k^{\rm uocc}G^+_k(\vr,\vr',\w)
\ee
where we have defined
\be
G_k^\pm(\vr,\vr',\w)=\frac{\vf_k(\vr)\vf_k(\vr')}{\w-\ve_k\pm i\eta}
\ee
with $\vf_k$ being a KS spin-orbital and $\ve_k$ the corresponding eigenvalue. 
Consider now a stretched homonuclear diatomic molecule composed of atom A and atom B. For large but finite inter-atomic separation $R$ the molecular KS orbitals can approximately be written as symmetric and antisymmetric linear combination of the atomic KS orbitals
\be
\vf_k(\vr)=\frac{1}{\sqrt{2}}[\vf^A_k(\vr)\pm\vf^B_k(\vr)]
\ee
where $\vf^A_k(\vr)$ is a KS orbital of atom A. This expression becomes exact as $R\to\infty$. It is easy to show that in the dissociation limit 
\be
E^{\rm RPA}[G_s]\to E^{\rm RPA}[G_s^A]+E^{\rm RPA}[G_s^B]
\label{elimrpa}
\ee
where
\bea
G_s^A&=&\sum_{k\ne 0}^{\rm occ}G^{-}_k+\sum_{k\ne 0}^{\rm uocc}G^{+}_k+\frac{1}{2}\left[G^{-}_0+G^{+}_0\right].
\eea
Here, $G_s^A$ contains only states of the isolated atom $A$. The orbital $k=0$ is the special orbital of the highest occupied state which has to be considered partially occupied and partially unoccupied. For a homoatomic system the fraction is always $1/2$. 

In the case of a covalent bonded heteronuclear diatomic molecule a similar analysis can be carried out. The only difference is that now only the highest occupied and lowest unoccupied arise from a degeneracy of the isolated atoms. Due to the lack of symmetry we also have to allow for a more general linear combination of KS orbitals
\bea
\vf_0^{\rm LUMO}(\vr)&=&\frac{1}{\sqrt{2}}[\sqrt{p}\vf^A_0(\vr)- \sqrt{(2-p)}\vf^B_0(\vr)]\\
\vf_0^{\rm HOMO}(\vr)&=&\frac{1}{\sqrt{2}}[\sqrt{(2-p)}\vf^A_0(\vr)+\sqrt{p} \vf^B_0(\vr)],
\eea
where $p\in [0,2]$. The KS Green functions to be inserted in Eq. (\ref{elimrpa}) become 
\bea
\label{greendiss1}
\!\!\!\!G_s^A&=&\sum_{k\ne 0}^{\rm occ}G^{-}_k+\sum_{k\ne 0}^{\rm uocc}G^{+}_k+\frac{p}{2}\,G^{-}_0+\frac{2-p}{2}G^{+}_0\\
\label{greendiss2}
\!\!\!\!G_s^B&=&\sum_{k\ne 0}^{\rm occ}G^{-}_k+\sum_{k\ne 0}^{\rm uocc}G^{+}_k+\frac{2-p}{2}G^{-}_0+\frac{p}{2}\,G^{+}_0.
\eea
In the exact system the energy assumes its minimum at $p=1$. This is, however, not guaranteed using an approximate functional. In fact, most functionals yield a so-called fractional charge error on the atomic fragments (i.e. the minimum is found at $p \ne 1$).
 
 \subsection{Fractional charge}
In this section we derive an expression for the RPA correlation energy in terms of  fractionally charged spin-compensated systems. To this end, we propose ensembles of the following form
\bea
\label{ens1}
\!\!\!\!\!\!\!\!\!\!\hat{\g}^<&=&(1-p)|0\,\ket\bra\, 0|+\frac{p}{2}(|\uparrow\,\ket\bra\, \uparrow|+|\downarrow\,\ket\bra\, \downarrow|)\\
\!\!\!\!\!\!\!\!\!\!&&{\rm for}\,\,p\in [0,1)\nn\\
\label{ens2}
\!\!\!\!\!\!\!\!\!\!\hat{\g}^>&=&\frac{2-p}{2}(|\uparrow\,\ket\bra\, \uparrow|+|\downarrow\,\ket\bra\, \downarrow|)+(p-1)|\uparrow\downarrow\,\ket\bra\, \uparrow\downarrow|\\
\!\!\!\!\!\!\!\!\!\!&&{\rm for}\,\,p\in (1,2]\nn
\eea
where $|0\ket$ refers to the ground state of $N_0$ even number of electrons, $|\! \uparrow\downarrow\ket$ referes to a spin-compensated $N_0+2$-state and $|\!\downarrow\ket$($|\!\uparrow\ket$) a spin-down(up) polarized state with $N_0+1$ electrons. The definition of the non-interacting spin-up $G^{\uparrow}_{\alpha}$ and spin-down $G^{\downarrow}_{\alpha}$ Green functions are
\bea
G^{\uparrow}_{\alpha}(\vr t,\vr' t')=\bra \,\alpha|T\{\psi^{\uparrow}(\vr t)\psi^{\uparrow}(\vr't')^{\dagger}\}|\alpha\,\ket\\
G^{\downarrow}_{\alpha}(\vr t,\vr' t')=\bra \,\alpha|T\{\psi^{\downarrow}(\vr t)\psi^{\downarrow}(\vr't')^{\dagger}\}|\alpha\,\ket
\eea
where $\alpha$ can be any of the states $|0\ket$, $|\!\downarrow\ket$, $|\!\uparrow\ket$ or $|\! \uparrow\downarrow\ket$ in the KS system. Here, $T$ is the time-ordering operator and $\psi^\b(\vr t)$ the field operator with the property of adding $\psi^{\b}(\vr t)^{\dagger}$ or removing $\psi^{\b}(\vr t)$ an electron with spin $\b$ in $\vr$ at time $t$. Evaluating the Green functions as ensemble expectation values using Eqs. (\ref{ens1}-\ref{ens2})
we find Green functions exactly of the form as in Eq. (\ref{greendiss1}-\ref{greendiss2}). Thus, the ensembles proposed correspond to how an atom in the RPA is described in the dissociation limit. 

The RPA correlation energy is usually evaluated using the time-ordered KS response function $\chi_s$. Using the ensemble Green functions we find
\bea
\chi_s^{E}(\vr,\vr',\w)&=&-i\pi\d(\w)p(2-p)|\vf_{0}(\vr)|^2|\vf_{0}(\vr')|^2\nn\\
&&+\tilde\chi_s^{p}(\vr,\vr',\w)
\label{chise}
\eea
where
\bea
\!\!\!\!\!\!\!\!\tilde\chi_s^{p}(\vr,\vr',\w)&=&2\sum_{q}\frac{f_q(\vr)f_q(\vr')}{\w-\ve_q+2i\eta}-\frac{f_q(\vr)f_q(\vr')}{\w+\ve_{q}-2i\eta}
\label{chitilde}
\eea
and we have performed a spin summation. The orbitals $\vf_k$ are now referring to the orbitals of an isolated atom. The index $q=(k,k')$ is a composite index of occupied ($k$) and unoccupied ($k'$) states. The special transition $q=(0,0)$ is excluded since it has been taken out explicitly (first term in Eq. (\ref{chise})). The functions $f_q(\vr)=\vf_k(\vr)\vf_{k'}(\vr)$ are called excitation functions and we note that when $\vf_0$ is occupied it should be multiplied by $\sqrt{p/2}$ and when it is unoccupied by $\sqrt{(2-p)/2}$. When calculating the correlation energy the interacting RPA response function has to first be evaluated $\chi_\l^E=\chi_s^E+\l\chi_s^Ev\chi_\l^E$. The final expression of the RPA correlation energy reads
\bea
\!\!\!\!\!\!\!E_{\rm c}&=&-\frac{p(2-p)}{4}\int d\vr d\vr'|\vf_{0}(\vr)|^2v(\vr,\vr')|\vf_{0}(\vr')|^2\nn\\
\!\!\!\!\!\!\!&&+\frac{i}{2}\int_0^1 d\l\int_{-\infty}^{\infty}\frac{d \w}{2\pi}\,\Tr \{v[\tilde\chi^p_\l(\w)-\tilde\chi^p_s(\w)]\}
\label{fracen}
\eea
where $\tilde\chi^p_\l(\w)$ is the scaled interacting response function calculated using Eq. (\ref{chitilde}). This equation will allow us to investigate the fractional spin error in the following section and to obtain numerical results for the fractional charge error in Section~\ref{results}.
\subsection{Fractional spin error}
The fractional spin error is defined as the energy difference of a system with $N_0+1$ electrons with one electron spin polarized (sp) on one hand and the same system but fully spin compensated (sc) on the other hand. Let us split the RPA interaction energy into the sum of Hartree and exchange energy (Hx) and the correlation energy. The sum of Hx energy in the sp case reads
\be
E^{\rm Hx}_{\rm sp} = \sum_{i,j}^{N_0/2} \langle ij||ij \rangle
+ \sum_{i}^{N_0/2} \langle 0i||0i \rangle
\ee
where we introduced the anti-symmetrized integral $\langle ij||kl \rangle = \int d\vr d\vr'\vf_{i}^{*}(\vr)\vf_{j}^{*}(\vr')v(\vr,\vr') \left ( 2\vf_{k}(\vr)\vf_{l}(\vr') - \vf_{l}(\vr)\vf_{k}(\vr') \right )$.
The second term accounts for all interactions of the singly occupied atomic orbital $\vf_0$. The expression for the spin compensated case reads
\begin{align}
E^{\rm Hx}_{\rm sc} = E^{\rm Hx}_{\rm sp}
%&\sum_{i,j}^{N/2} 2 \langle ij||ij \rangle  + \sum_{i}^{N/2} \langle 0i||0i \rangle \nn \\
+ \frac{1}{4}\int d\vr d\vr'|\vf_{0}(\vr)|^2v(\vr,\vr')|\vf_{0}(\vr')|^2.
\end{align}
The additional term is a nonzero self-interaction term, which is equal to the fractional spin error of EXX. It is this term alone that is responsible for the wrong dissociation limit of EXX for H$_{2}$. It also introduces a large additional error in the dissociation limit of any system, on top of the error inherent in the EXX functional when describing the fragments.

We now turn to the RPA correlation energy. The correlation energy for the sc system is taken from Eq.~\eqref{fracen} with $p=1$.
\bea
\!\!\!\!\!\!\!E_{\rm sc}^{\rm c}&=&-\frac{1}{4} \int d\vr d\vr'|\vf_{0}(\vr)|^2v(\vr,\vr')|\vf_{0}(\vr')|^2\nn\\
\!\!\!\!\!\!\!&&+\frac{i}{2}\int_0^1 d\l\int_{-\infty}^{\infty}\frac{d \w}{2\pi}\,\Tr \{v[\tilde\chi^1_\l(\w)-\tilde\chi^1_s(\w)]\}
\label{fracen-p1}
\eea
We now see that the first term exactly cancels the fractional spin error of EXX. The second term in Eq. (\ref{fracen}) is identical to the correlation energy obtained in the spin-polarized case. Consequently, the RPA functional does not suffer from fractional spin error. There will, however, still be a self-correlation due to the second term but this is, in general, expected to be smaller. These conclusions are confirmed by the numerical results obtained previously\cite{nedvlvb2,fngb05,nedvlvb} and here in Sec. V.
\section{Computational details}
Our goal is to find the local potential that minimizes the RPA energy functional. For the three dimensional (3D) calculations we utilize the direct minimization scheme for the optimized effective
potential (OEP) proposed by Yang and Wu.~\cite{yang} The potential is expanded in a basis plus a reference potential
\be
v(\br) = v_{0}(\br) + \sum_{t} b_{t} g_{t}(\br).
\ee
As a reference potential we use the sum of the nuclear potential and the Fermi-Amaldi potential, $v^{FA}(\br)$, evaluated with the Hartree-Fock density
\be
v^{FA}(\br) = \frac{N-1}{N} \int \frac{n(\br')}{|\br - \br'|} d\br'.
\ee
For closed shell systems the derivative of the total energy functional with respect to the expansion coefficients, $b_{t}$, is readily evaluated via Eq.~\eqref{derivative}. For systems with fractional charge we evaluated the gradient with the finite difference method.
The gradient is then fed to the Broyden-Fletcher-Goldfarb-Shanno optimization routine to find the minimum total energy. The calculations
were considered converged when the vector norm of the gradient was less than $10^{-3}$.
The algorithm was implemented in a local version of the PyQuante module.~\cite{pyquante}

It is a well-known fact that finite basis set OEP calculations can become numerically unstable if a too large potential basis set is used.~\cite{Rohr1,Rohr2,
Hesselmann1} We carefully chose
the potential basis sets to be balanced to the respective orbital basis sets. The proper choice is reflected in the smooth potentials that we obtain. For our calculations we used the standard orbital basis cc-pVTZ from the Dunning family. 

For the potential bases we used even tempered gaussians. Each set is characterized by three numbers: the smallest exponent, $\a_{\rm min}$, the largest exponent, $\a_{\rm max}$, and the number of basis functions, $N$.
With this set of parameters we construct the exponent $\a_i$ as 
\be
\label{potbas}
\a_i=\a_{\rm min}\left[\frac{\a_{\rm max}}{\a_{\rm min}}\right]^{\left(\frac{i-1}{N-1}\right)},
\ee 
where $i$ runs from 1 to $N$. For the atomic calculations we use only s-type functions. For molecular calculations we add a set of p-type functions using the same exponents and omitting the largest.

For the one dimensional (1D) calculations we use a basis set of cubic splines, which permits us to solve Eq. (\ref{lss}) by inverting the static KS response function. The spline-basis is described in detail in several previous works where it has been used successfully for OEP-type of equations.\cite{splines,hvb1,hvB3}
\section{Results}
\label{results}
As a first step we verify our implementation. To this end, we compare the total energy and the correlation potential with accurate reference data for He, Be, and Ne.\cite{hvb4} The total energies and ionization potentials (IP) are listed in table~\ref{atoms}. The potential basis sets are given by $\a_{\rm min} = 0.01$, $\a_{\rm max} = 10$, and $N=5$ for He, $\a_{\rm min} = 0.01$, $\a_{\rm max} = 10$, and $N=9$ for Ne, and $\a_{\rm min} = 0.001$, $\a_{\rm max} = 100$, and $N=11$ for Ne (compare equation~\eqref{potbas}).
\begin{table}[htbp]
   \centering
   \begin{tabular}{l|rr|rr} % Column formatting, @{} suppresses leading/trailing space
        &\multicolumn{2}{c|}{this work} & \multicolumn{2}{c}{benchmark} \\
        Atom & IP & energy &  IP & energy \\
        \hline
        He & -0.907 & -2.936 & -0.902  & -2.945 \\
        Be & -0.349 & -14.681 & -0.354 & -14.754 \\
        Ne & -0.773 & -128.945 & -0.796  & -129.103
        \end{tabular}
   \caption{Energies in Hartree.}
   \label{atoms}
\end{table}

\begin{figure}[t] %  figure placement: here, top, bottom, or page
   \centering
   \includegraphics[width=3.5in]{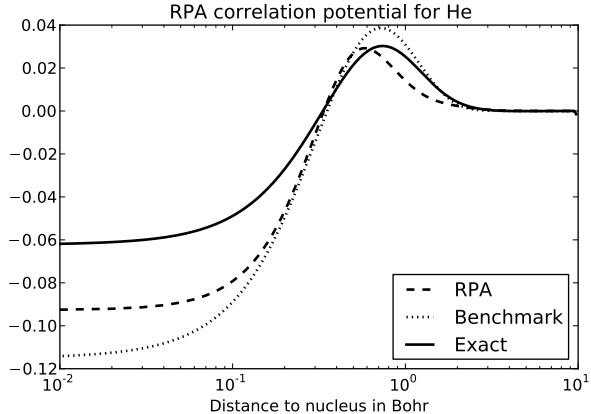} 
   \caption{The correlation potential of He for RPA\cite{hvb4} (dotted), for RPA in this work (dashed) and the exact correlation potential\cite{umr} (solid).}
   \label{fig:He}
\end{figure}

\begin{figure}[t] %  figure placement: here, top, bottom, or page
   \centering
   \includegraphics[width=3.5in]{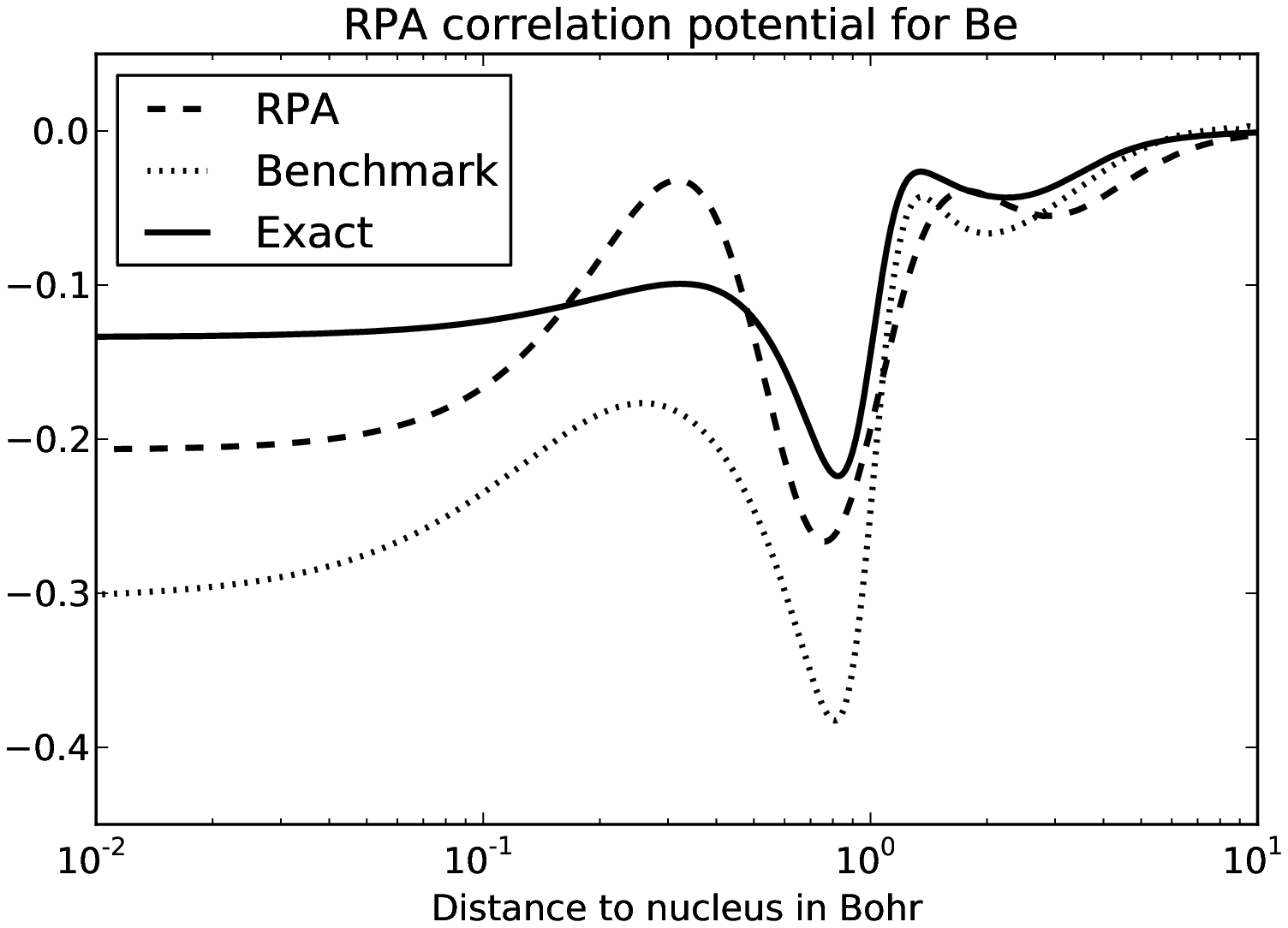} 
   \caption{The correlation potential of Be for RPA\cite{hvb4} (dotted), for RPA in this work (dashed), and the exact correlation potential\cite{umr} (solid).}
   \label{fig:Be}
\end{figure}

\begin{figure}[t] %  figure placement: here, top, bottom, or page
   \centering
   \includegraphics[width=3.5in]{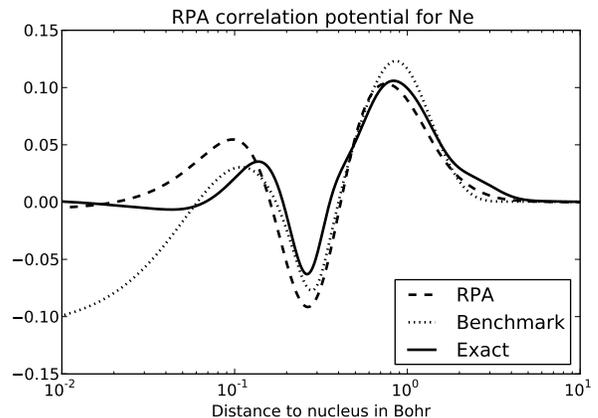} 
   \caption{The correlation potential of Ne for RPA\cite{hvb4} (dotted), for RPA in this work (dashed), and the exact correlation potential\cite{umr} (solid).}
   \label{fig:Ne}
\end{figure}

The total energies of our new implementation are somewhat higher than the reference energies. The differences range from 9 mH to 158 mH. This result is expected since the reference calculations were performed with a virtually complete orbital basis. The differences in IP are minor (from 5 mH to 23 mH), thus indicating an accurate potential. This conclusion is supported by figures~\ref{fig:He} -~\ref{fig:Ne}, which show our RPA correlation potentials (dashed) of He, Be, and Ne comparing
to accurate RPA correlation potentials (dotted).\cite{hvb4} The RPA correlation potentials shown in the figures are calculated as
\be
\label{corpot}
v_c^{\rm RPA}(\vr)=v^{\rm RPA}_{\rm KS}[n_{\rm RPA}](\vr)-v^{\rm EXX}_{\rm KS}[n_{\rm EXX}](\vr).
\ee
In all figures we qualitatively reproduce the reference potentials. The largest deviation is found close to the nucleus (the x-axis is on a logarithmic scale). For comparison we also show the exact KS correlation potential (solid)\cite{umr} and we see that the RPA correlation potential closely resembles the exact correlation potential.

\begin{figure}[t] %  figure placement: here, top, bottom, or page
   \centering
   \includegraphics[width=3.5in]{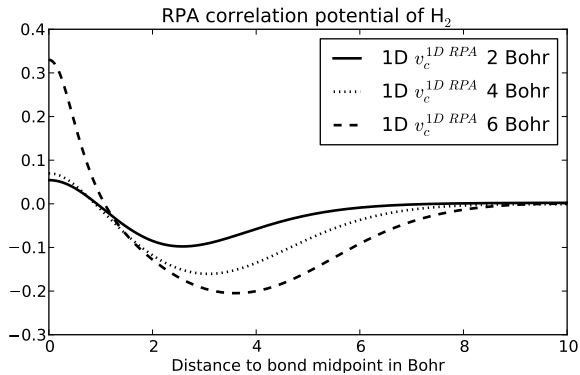} 
   \caption{The correlation potential along the bond axis with the bond midpoint at zero. The system is 1D H$_{2}$ with a soft coulomb potential.}
   \label{fig:h2pot1D}
\end{figure}

To analyze the RPA molecular correlation potential we first investigate a 1D model system with soft coulomb interactions. The nuclear potentials are $-Z/\sqrt{(x\pm R/2)^2+1}$, where $Z$ is the nuclear charge and $R$ is the internuclear distance. The electron-electron interaction is set to $1/\sqrt{(x_1-x_2)^2+1}$. The first system consists of two H atoms ($Z=1.2$). Figure~\ref{fig:h2pot1D} shows the RPA correlation potentials along the bond axis at bond distances of 2, 4, and 6 Bohr, with the bond midpoint at zero. We notice that the minimum of the correlation potential is shifted away from the atom, but that the shift decreases with increasing interatomic distance. At the bond midpoint a positive peak emerges with increasing bond distance. Both features are also observed for the exact KS correlation potential.\cite{tmm09,gb96}

\begin{figure}[t] %  figure placement: here, top, bottom, or page
   \centering
   \includegraphics[width=3.5in]{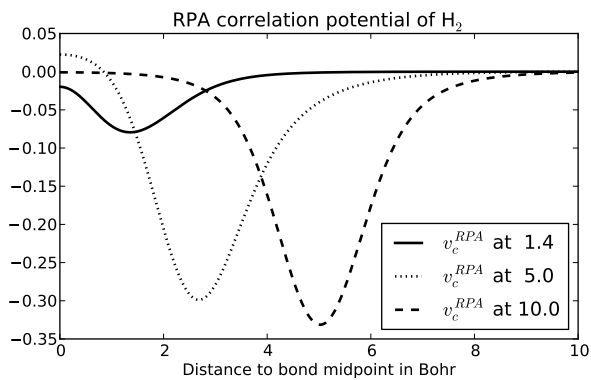} 
   \caption{The correlation potential of H$_{2}$ along the bond axis. The bond midpoint is at zero. We show the RPA correlation potential
   for interatomic distance 1.4 (solid), 5.0 (dotted), and 10.0 (dashed).}
   \label{fig:h2pot}
\end{figure}

With our new implementation we are able to investigate the H$_{2}$ molecule in all three dimensions and with the full coulomb interaction (potential basis: $\a_{\rm min} = 0.1$, $\a_{\rm max}=1.0$, and $N=5$).
In Figure~\ref{fig:h2pot} we show the RPA correlation potential (see equation~\eqref{corpot}) along the bond axis, where again the bond midpoint is at zero. We show the potentials for equilibrium distance (1.4 Bohr; solid curve), for 5.0 (dotted), and for 10.0 Bohr (dashed). The potentials qualitatively resemble the potentials obtained from the 1D calculations (c.f.~\ref{fig:h2pot1D}). Only the peak at the bond midpoint is absent for small (1.4) and large (10.0) interatomic separation. For large separation the absence is easily explained. The orbital and potential basis functions are simply not diffuse enough to extend to the bond midpoint. This results in a vanishing correlation potential at the bond midpoint.
The situation is different for a bond distance of 1.4 Bohr. With a small atomic separation the orbital and potential basis extend to the bond midpoint as is evident from the non-vanishing correlation potential. However, the potential basis functions are not compact enough to produce a peak. We have verified this by using more compact potential basis functions. With this set of basis functions, however, we find unphysical oscillations for larger atomic separations. We also placed orbital and potential functions at various points along the bond axis. We observed a peak in the potential at each of the points, which leads us to conclude that these peaks are artifacts of the basis rather than genuine features of the functional.

At this point we restrain from showing the dissociation curve and rather point the reader to a future publication with a detailed discussion of the dissociation curves. We would only like to mention that the well-known "bump"\cite{nedvlvb2,fngb05,nedvlvb,hessg11} is still present, but somewhat weaker.


\begin{figure}[t] %  figure placement: here, top, bottom, or page
   \centering
   \includegraphics[width=3.5in]{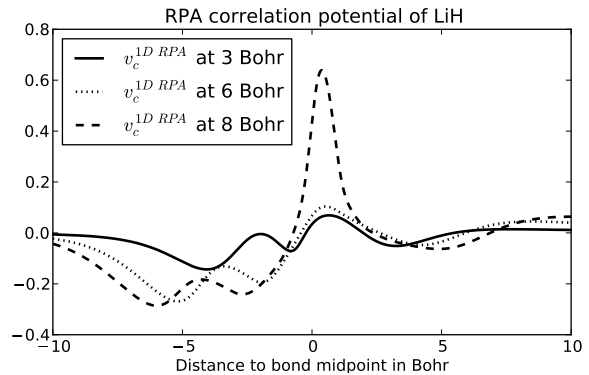} 
   \caption{The correlation potential along the bond with the bond midpoint at zero. The system is 1D LiH with a soft coulomb potential. The Li atom is located at -1, -1.5, and -4 Bohr, respectively. The H atom is located at 1, 1.5, and 4 Bohr, respectively.}
   \label{fig:lihpot1D}
\end{figure}

We now turn to the LiH molecule. In figure~\ref{fig:lihpot1D} we show the RPA correlation potential (defined in equation~\eqref{corpot}) of 1D LiH ($Z_{\rm Li}=3.6,Z_{\rm H}=1.2$). The
solid, dotted and dashed curves represent the correlation potentials for 2, 3, and 8 Bohr interatomic distances, respectively. The build up of the peak at the bond midpoint is apparent. However, a step, as is present in the exact KS correlation potential, is not observed.

\begin{figure}[t] %  figure placement: here, top, bottom, or page
   \centering
   \includegraphics[width=3.5in]{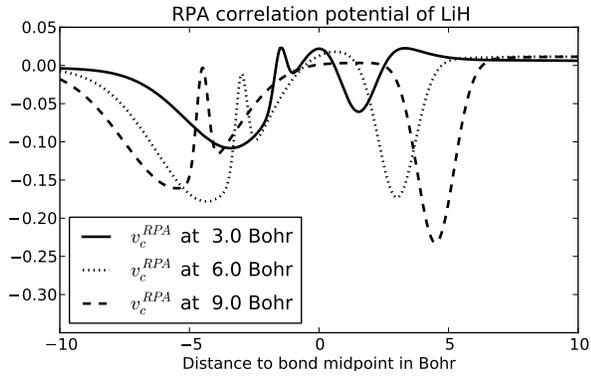} 
   \caption{The correlation potential of LiH along the bond axis. The bond midpoint is at zero. The Li atom is located at -1.5, -2.5, and -4 Bohr, respectively. The H atom is located at 1.5, 2.5, and 4 Bohr, respectively.}
   \label{fig:lihpot}
\end{figure}

Figure~\ref{fig:lihpot} shows the correlation potential (equation~\eqref{corpot}) of the three dimensional LiH for bond distances 3 (solid), 6 (dotted), and 8 Bohr (dashed). The parameters for the potential basis of the H atom are $\a_{\rm min} = 0.01$, $\a_{\rm max} = 1.0$, and $N=2$. For the Li atom they read $\a_{\rm min} = 0.001$, $\a_{\rm max} = 10$, and $N=3$.
The same features as in the 1D case are also found in the results for the 3D correlation potentials. In the region of the Li atom (-10 to 0) the potential qualitatively 
resembles that of the 1D system in Figure~\ref{fig:lihpot1D}. We see a well with a peak close to the nucleus. Also in the region of the H atom (0 to 10)
figures~\ref{fig:lihpot1D} and~\ref{fig:lihpot} show the same structure. A well emerges with increasing bond distance.
The difference between 1D and 3D is found only at the bond midpoint. Like in the case of H$_{2}$ a peak emerges only for the 1D system (figure~\ref{fig:lihpot1D}). In contrast, the 3D system (figure~\ref{fig:lihpot}) exhibits a peak at zero only for small and intermediate bond distances (3 and 5 Bohr). The explanation for the absence of the peak at the bond midpoint for 8 Bohr is the same as in the case of H$_{2}$.

\begin{figure}[t] %  figure placement: here, top, bottom, or page
   \centering
   \includegraphics[width=3.5in]{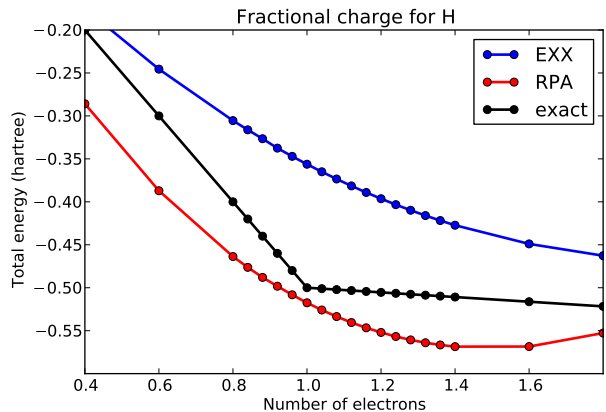} 
   \caption{The total energy of H as a function of number of electrons for EXX (blue), RPA (red), and exact (black).}
   \label{fig:H}
\end{figure}

\begin{figure}[t] %  figure placement: here, top, bottom, or page
   \centering
   \includegraphics[width=3.5in]{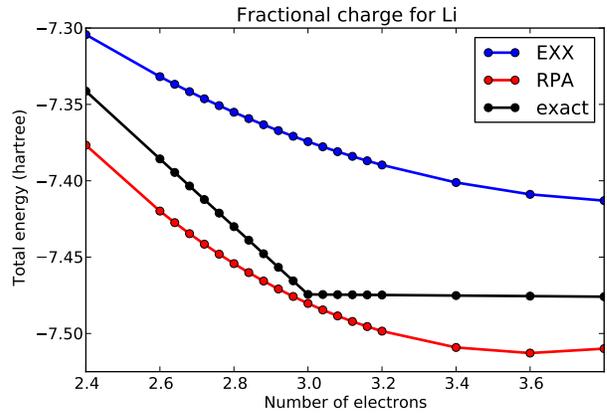} 
   \caption{The total energy of Li as a function of number of electrons for EXX (blue), RPA (red), and exact (black).}
   \label{fig:Li}
\end{figure}

\begin{figure}[t] %  figure placement: here, top, bottom, or page
   \centering
   \includegraphics[width=3.5in]{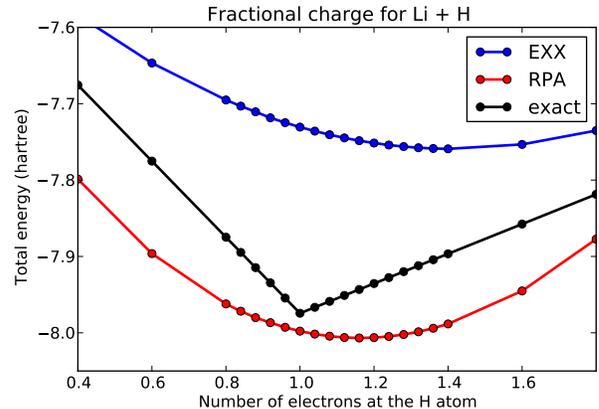} 
   \caption{The total energy of dissociated LiH as a function of number of electrons at the H atom for EXX (blue), RPA (red), and exact (black).}
   \label{fig:LiHfrac}
\end{figure}

We further analyze the RPA functional in the context of fractional charge.  The total energy of H and Li (potential bases like in LiH) is plotted (figures~\ref{fig:H} and~\ref{fig:Li}) as a function of the number of electrons, where we allow fractional values, according to Eq. \ref{fracen}. Both atoms show a smooth behavior, thus showing that the RPA (red) misses the kink at integer electron numbers. Comparing to the exact curves we see, however, a large improvement compared to the EXX (blue) functional regarding the total energy. The agreement is particularly well at integer number of electrons. This explains the good dissociation energies for homoatomic systems found in the RPA.
We can also combine these two figures to analyze the RPA energy of LiH in the dissociation limit. In order to do this we add the energies for the H atom to the energies of the Li atom so that the total number of electrons sums up to four. Figure~\ref{fig:LiHfrac} shows the total energy as a function of the number of electrons at the H atom. The number of electrons at the Li atom will then be four minus the x value.
The exact functional (black) has a minimum at 1.0, because it dissociates LiH into a neutral H atom and a neutral Li atom. In Figure~\ref{fig:LiHfrac} we see that EXX (blue) and RPA (red) do not dissociate LiH into the neutral atoms. In both cases there is a surplus of electronic charge at the H atom. However, in the case of RPA the surplus (0.16 electrons) is much smaller than in the case of EXX (0.4 electrons). The large improvement may be related to the peak that is present in the RPA correlation potential. 

Finally, in table~\ref{molecules} we collect the total energies in the dissociation limit of H$_2$, Li$_2$, and LiH for the exact functional, EXX, and RPA. Please note that the LiH energies of EXX and RPA are taken as the minimum of the respective curves in figure~\ref{fig:LiHfrac} and not the values at 1.0. This is to account for the fractional charge error present in both functionals. 
\begin{table}[htbp]
   \centering
   %\topcaption{Table captions are better up top} % requires the topcapt package
   \begin{tabular}{l|rrr} % Column formatting, @{} suppresses leading/trailing space
%        &\multicolumn{2}{c|}{this work} & \multicolumn{2}{c}{benchmark} \\
         & exact & EXX &  RPA \\
        \hline
        H$_2$ & -1.000 & -0.713 & -1.035 \\
        Li$_2$ & -14.948 & -14.749 & -14.961 \\
        LiH & -7.974 & -7.759 & -8.007
        \end{tabular}
   \caption{Total energies in Hartree in the dissociation limit.}
   \label{molecules}
\end{table}

\section{Conclusions}
We have obtained self-consistent RPA correlation potentials for diatomic molecules and studied their behavior as a function of interatomic distances. At large distances the RPA potential correctly exhibits a peak at the bond midpoint but misses the step feature. 

We have also analyzed the RPA functional for fractional charges by evaluating the ensemble averaged KS Green function using spin-compensated ensembles with non-integer number of particles. This procedure can easily be carried over to any functional constructed from the KS Green function.  
 
The numerical results show that the kink at integer number of electrons is missed in the RPA. As a consequence, we find spurious fractional charges on the dissociated fragments. The charges are, however, much smaller compared to other functionals. On the other hand, the RPA will most likely not be able to describe the so-called field counteracting effect in the correlation potential of hydrogen chains which has been discussed to have its origin in the derivative discontinuity.\cite{fieldca}

In summary, we have found that the RPA accomplishes the following:
\begin{itemize}
\item The dissociation limit of closed-shell molecules is well reproduced in the RPA, due to the explicit elimination of the self-interaction term present in the EXX functional.
\item RPA separates the charges in bond dissociation by virtue of a peak at the bond midpoint, an exact feature of the true correlation potential. 
\item RPA exhibits only a small fractional charge error in the cases of closed-shell covalently bonded molecules. 
\end{itemize}
These findings consolidate the high expectations on the RPA that are currently prevalent. There is, however, still room for improvement as the discontinuity at odd integer particle number is missing. We believe that improvements on the RPA within the ACFD framework will help to overcome this shortcoming. 

\appendix
\section{Derivative of the RPA energy}
In this Appendix we return to Eq. (\ref{lss}) and evaluate $n_c(\vr)$ in terms of KS orbitals, $\vf_k$, and KS eigenvalues, $\ve_k$. The self-energy (Eq. (\ref{sigma})) involves an integration over the frequency of the following form
\begin{equation}
\S_c(\w)=i\int \frac{d\omega'}{2\pi}G_s(\omega'+\w)v\chi^{\rm RPA}(\omega')v,
\label{sigmarpa}
\end{equation}
where we have suppressed the space-coordinates. To perform the integration we write the KS Green function in its Lehmann representation
\bea
G_s(\vr,\vr',\w)&=&\sum_k\vf_k(\vr)\vf_k(\vr')\nn\\
&&\times\left[\frac{n_k}{\w-\ve_k-i\eta}+\frac{1-n_k}{\w-\ve_k+i\eta}\right],
\eea
where $n_k$ is the occupation number. The response function in the RPA is given by Eq. (\ref{rpa}). This equation can easily be rewritten as an eigenvalue problem in terms of the matrix
\bea
V_{qq'}=\w_{q}^2\d_{qq'}+\bra \tilde {f}_q|v|\tilde {f}_{q'}\ket,
\eea 
where $\tilde{f}_q(\vr)=2\sqrt{\w_q}f_q(\vr)$, with $f_q(\vr)$ being a KS excitation function and $\w_q$ a KS excitation energy. The square root of the eigenvalues corresponds to the true excitation energies $Z_q$ and the excitation functions are transformed according to 
\begin{equation}
F_{q}(\vr)=\sum_{q'}f_{q'}(\vr)U_{q'q},
\end{equation}
where $U$ is the matrix diagonalizing $V$. The interacting time-ordered response function in the RPA can then be written as
\bea
\chi^{\rm RPA}(\vr,\vr',\w)&=&\sum_q\frac{1}{2Z_q}F_q(\vr)F_q(\vr')\nn\\
&&\times\left[\frac{1}{\w-Z_q+i\eta}-\frac{1}{\w+Z_q-i\eta}\right]
\eea
and the integral in Eq. (\ref{sigmarpa}) becomes
\bea
\S_c(\vr,\vr',\w)=\sum_{kq}\vf_k(\vr)\int d\vr_1v(\vr,\vr_1)F_q(\vr_1)\nn\\
\times\vf_k(\vr')\int d\vr_1v(\vr',\vr_1)F_q(\vr_1)\nn\\
\times\frac{1}{2Z_q}\left[\frac{1-n_k}{\omega-\varepsilon_k-Z_{q}+
i{\eta}}+\frac{n_k}{\omega-\varepsilon_k+Z_{q}-
i{\eta}}\right].
\eea
Next, we evaluate
\bea
n_c=-i\int \frac{d\w}{2\pi}\S_c(\w)G_s(\w)G_s(\w).
\eea
After performing standard contour integrations we get in total six terms, which after symmetry considerations can be reduced to four. In summary, we find
\begin{eqnarray}\label{ncs}
\!\!\!\!\!n_c(\vr)&=&\sum_{ksp}\sum_{q}\frac{1}{Z_q}\left(\ldots\right)\varphi^{*}_s(\vr)\varphi_p(\vr) \nonumber\\
\!\!\!\!\!&\times&\int d\vr_1d\vr_2F_q(\vr_1)v(\vr_1,\vr_2)\varphi_k(\vr_2)\varphi^{*}_p(\vr_2)\nonumber\\
\!\!\!\!\!&\times&\int d\vr_1d\vr_2\varphi_s(\vr_1)\varphi^{*}_k(\vr_1)v(\vr_1,\vr_2)F^{*}_q(\vr_2).
\end{eqnarray}
where the dots signifies the insertion of the following four terms
\begin{eqnarray}
-\frac{(1-n_k)n_pn_s}{(\ve_k+Z_{q}-\ve_p)(\ve_k+Z_{q}-\ve_s)}\\
+\frac{2(1-n_k)n_p(1-n_s)}{(\ve_p-Z_{q}-\ve_k)(\ve_p-\ve_s)}\\
+\frac{2n_kn_p(1-n_s)}{(\ve_s+Z_{q}-\ve_k)(\ve_p-\ve_s)}\\+\frac{n_k(1-n_s)(1-n_p)}{(\ve_s+Z_{q}
-\ve_k)(\ve_p+Z_{q}-\ve_k)}.
\end{eqnarray}
The expression in Eq. (\ref{ncs}) has, in this work, been implemented both in the 3D and in the 1D case.
%------------------------------------------------------------------------------------------------------------
%---------------------------------------------------C&D--------------------------------------------------
%------------------------------------------------------------------------------------------------------------
 %\begin{acknowledgments}
%This work was supported by...
%\end{acknowledgments}
%\bibliography{../refrpa}

\end{document}